\documentclass[twocolumn,A4]{article}
\usepackage[dvips]{graphics}
\usepackage{setspace}
\usepackage{color}
\definecolor{gold}{rgb}{0.85,0.66,0}
\definecolor{dblue}{rgb}{0,0,0.8}
\begin{document}
\onecolumn
\begin{center}
{\bf{\Large {\textcolor{gold}{A mesoscopic ring as a XNOR gate: An 
exact result}}}}\\
~\\
{\textcolor{dblue}{Santanu K. Maiti}}$^{1,2,*}$ \\
~\\
{\em $^1$Theoretical Condensed Matter Physics Division,
Saha Institute of Nuclear Physics, \\
1/AF, Bidhannagar, Kolkata-700 064, India \\
$^2$Department of Physics, Narasinha Dutt College,
129, Belilious Road, Howrah-711 101, India} \\
~\\
{\bf Abstract}
\end{center}
We describe XNOR gate response in a mesoscopic ring threaded by a magnetic 
flux $\phi$. The ring is attached symmetrically to two semi-infinite 
one-dimensional metallic electrodes and two gate voltages, viz, $V_a$ and
$V_b$, are applied in one arm of the ring which are treated as the inputs 
of the XNOR gate. The calculations are based on the tight-binding model 
and the Green's function method, which numerically compute the 
conductance-energy and current-voltage characteristics as functions of 
the ring-to-electrode coupling strength, magnetic flux and gate voltages. 
Our theoretical study shows that, for a particular value of $\phi$ 
($=\phi_0/2$) ($\phi_0=ch/e$, the elementary flux-quantum), a high output 
current ($1$) (in the logical sense) appears if both the two inputs to 
the gate are the same, while if one but not both inputs are high ($1$), 
a low output current ($0$) results. It clearly exhibits the XNOR gate 
behavior and this aspect may be utilized in designing an electronic 
logic gate. 

\vskip 1cm
\begin{flushleft}
\end{flushleft}
\vskip 4.2in
\noindent
{\bf ~$^*$Corresponding Author}: Santanu K. Maiti

Electronic mail: santanu.maiti@saha.ac.in

\newpage
\twocolumn

\section{Introduction}

Low-dimensional model quantum confined systems have been the key objects 
of both theoretical and experimental research, mainly due to the fact that 
these simple looking systems are prospective candidates for nanodevices 
in electronic as well as spintronic engineering. The study of electron 
transport in such systems has become one of the most fascinating branch 
of nanoscience and technology. The idea of manufacturing nanodevices are 
based on the concept of quantum interference effect. For much smaller 
sizes the quantum phase coherence is maintained across the sample, while 
it generally disappears for larger systems. A mesoscopic metallic ring 
is one such promising example where electronic motion is confined and 
the transport becomes predominantly coherent. With the help of a 
mesoscopic ring, we can make an electronic device that can be operated 
as a logic gate, which may be used in nanoelectronic circuits. To explore 
this phenomenon we design a bridge system where the ring is sandwiched 
between two external electrodes, the so-called electrode-ring-electrode 
bridge. The ring is then subjected to an Aharonov-Bohm (AB) flux $\phi$ 
which is the key controlling factor for the whole logical operation in 
this particular geometry. Following the pioneering work of Aviram and 
Ratner,$^1$ the theoretical description of electron transport 
in a bridge system has got much progress. Later, many excellent 
experiments$^{2-4}$ have been done in several bridge systems to 
understand the basic mechanisms underlying the electron transport. 
Though extensive studies on electron transport have already been done 
both theoretically$^{5-12}$ as well as experimentally,$^{2-4}$ yet 
lot of controversies are still present between the theory and experiment, 
and the complete knowledge of the conduction mechanism in this scale 
is not very well established even today. For illustrative purposes, here 
we mention some of these issues as follow. The electronic transport
in the ring changes drastically depending on the interface geometry 
between the ring and the electrodes. By changing the geometry, one 
can tune the transmission probability of an electron across the ring 
which is solely due to the effect of quantum interference among the 
electronic waves passing through different arms of the ring. Not only
that, the electron transport in the ring can be modulated in other 
way by tuning the magnetic flux, that threads the ring. The AB flux 
threading the ring may change the phases of the wave functions 
propagating along the different arms of the ring leading to constructive 
or destructive interferences, and therefore, the transmission amplitude 
changes.$^{13-17}$ Beside these factors, ring-to-electrode coupling is 
another important issue that controls the electron transport in a 
meaningful way.$^{17}$ All these are the key factors which regulate 
the electron transmission in the electrode-ring-electrode bridge 
system and they have to be taken into account properly to reveal 
the transport mechanisms. 

Our main aim of the present work is to explore the XNOR gate behavior in a 
mesoscopic ring threaded by a magnetic flux $\phi$. The ring is sandwiched 
symmetrically between two electrodes, and two gate voltages $V_a$ and $V_b$ 
are applied in one arm of the ring (see Fig.~\ref{xnor}) those are treated
as the two inputs of the XNOR gate. A simple tight-binding model is used 
to describe the system and all the calculations are done numerically. 
Here we address the XNOR gate behavior by studying the conductance-energy 
and current-voltage characteristics as functions of the ring-to-electrode 
coupling strength, magnetic flux and gate voltages. Our exact numerical
study reveals that for a particular value of the magnetic flux, 
$\phi=\phi_0/2$, a high output current ($1$) (in the logical sense) is 
available if both the two inputs to the gate are the same (either low 
or high), while if one but not both inputs are high ($1$), a low output 
current ($0$) appears. This phenomenon clearly shows the XNOR gate 
behavior. To the best of our knowledge the XNOR gate response in such a 
simple system has yet not been addressed in the literature.

We arrange the paper as follow. Following the introduction (Section $1$), 
in Section $2$, we described the model and the theoretical formulations 
for the calculation. Section $3$ explores the results, and finally, we 
conclude our study in Section $4$.

\section{Model and the theoretical background}

\begin{figure}[ht]
{\centering \resizebox*{7cm}{5cm}{\includegraphics{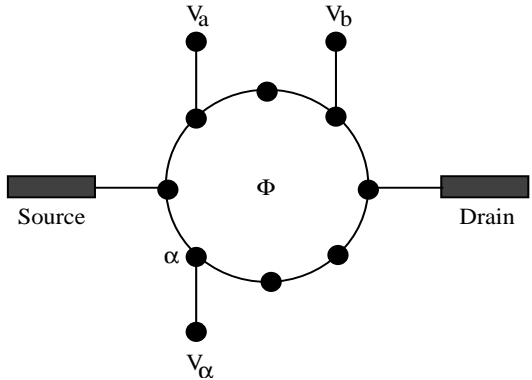}}\par}
\caption{Schematic view for the operation of a XNOR gate. The atomic 
sites $a$, $b$ and $\alpha$ are subjected to the voltages $V_a$, $V_b$ 
and $V_{\alpha}$, respectively, those are variable.}
\label{xnor}
\end{figure}

Let us start with the model presented in Fig.~\ref{xnor}. A mesoscopic ring, 
subjected to an AB flux $\phi$, is attached symmetrically (upper and lower 
arms have equal number of lattice points) to two semi-infinite 
one-dimensional ($1$D) metallic electrodes. Two gate voltages $V_a$ and
$V_b$, taken as the two inputs of the XNOR gate, are applied to the atomic 
sites $a$ and $b$, respectively, in the upper arm of the ring. While, an 
additional gate voltage $V_{\alpha}$ is applied to the site $\alpha$ in 
the lower arm of the ring. These three voltages are variable.

At much low temperatures and bias voltage, the conductance $g$ of the 
ring can be expressed from the Landauer conductance 
formula,$^{18,19}$
\begin{equation}
g=\frac{2e^2}{h} T
\label{equ1}
\end{equation}
where $T$ gives the transmission probability of an electron across 
the ring. This $(T)$ can be represented in terms of the Green's 
function of the ring and its coupling to the two electrodes by the 
relation,$^{18,19}$
\begin{equation}
T=Tr\left[\Gamma_S G_{R}^r \Gamma_D G_{R}^a\right]
\label{equ2}
\end{equation}
where $G_{R}^r$ and $G_{R}^a$ are respectively the retarded and 
advanced Green's functions of the ring including the effects of 
the electrodes. Here $\Gamma_S$ and $\Gamma_D$ describe the 
coupling of the ring to the source and drain, respectively. For 
the complete system i.e., the ring, source and drain, the Green's 
function is defined as,
\begin{equation}
G=\left(E-H\right)^{-1}
\label{equ3}
\end{equation}
where $E$ is the injecting energy of the source electron. To Evaluate
this Green's function, the inversion of an infinite matrix is needed since
the complete system consists of the finite ring and the two semi-infinite 
electrodes. However, the entire system can be partitioned into sub-matrices 
corresponding to the individual sub-systems and the Green's function for 
the ring can be effectively written as,
\begin{equation}
G_{R}=\left(E-H_{R}-\Sigma_S-\Sigma_D\right)^{-1}
\label{equ4}
\end{equation}
where $H_{R}$ is the Hamiltonian of the ring. Within the non-interacting 
picture the Hamiltonian can be expressed in the form,
\begin{eqnarray}
H_{R} & = & \sum_i \left(\epsilon_{i0} + V_a \delta_{ia} + V_b \delta_{ib} 
+ V_{\alpha} \delta_{i\alpha}
\right) c_i^{\dagger} c_i \nonumber \\
 & + & \sum_{<ij>} t \left(c_i^{\dagger} c_j e^{i\theta}+ c_j^{\dagger}
c_i e^{-i\theta}\right)
\label{equ5}
\end{eqnarray}
where $\epsilon_{i0}$'s are the site energies for all the sites $i$ 
except the sites $i=a$, $b$ and $\alpha$ where the gate voltages $V_a$, 
$V_b$ and $V_{\alpha}$ are applied, those are variable. These gate 
voltages can be 
incorporated through the site energies as expressed in the above 
Hamiltonian. $c_i^{\dagger}$ ($c_i$) is the creation (annihilation) 
operator of an electron at the site $i$ and $t$ is the nearest-neighbor 
hopping integral. The phase factor $\theta=2 \pi \phi/N \phi_0$ comes 
due to the flux $\phi$ threaded by the ring, where $N$ corresponds to 
the total number of atomic sites in the ring. A similar kind of 
tight-binding Hamiltonian is also used, except the phase factor 
$\theta$, to describe the $1$D perfect electrodes where the 
Hamiltonian is parametrized by constant on-site potential $\epsilon_0$ 
and nearest-neighbor hopping integral $t_0$. The hopping integral between
the source and the ring is $\tau_S$, while it is $\tau_D$ between the
ring and the drain. In Eq.~\ref{equ4}, $\Sigma_S=h_{SR}^{\dagger}g_S h_{SR}$
and $\Sigma_D=h_{DR} g_D h_{DR}^{\dagger}$ are the self-energy operators
due to the two electrodes, where $g_S$ and $g_D$ correspond to the Green's
functions of the source and drain, respectively. $h_{SR}$ and $h_{DR}$ are
the coupling matrices and they will be non-zero only for the adjacent
points of the ring, and the electrodes, respectively. The matrices
$\Gamma_S$ and $\Gamma_D$ can be calculated through the expression,
\begin{equation}
\Gamma_{S(D)}=i\left[\Sigma_{S(D)}^r-\Sigma_{S(D)}^a\right]
\label{equ6}
\end{equation}
where $\Sigma_{S(D)}^r$ and $\Sigma_{S(D)}^a$ are the retarded and advanced
self-energies, respectively, and they are conjugate with each other.
Datta {\em et al.}$^{20}$ have shown that the self-energies can be
expressed like as,
\begin{equation}
\Sigma_{S(D)}^r=\Lambda_{S(D)}-i \Delta_{S(D)}
\label{equ7}
\end{equation}
where $\Lambda_{S(D)}$ are the real parts of the self-energies which
correspond to the shift of the energy eigenvalues of the ring and
the imaginary parts $\Delta_{S(D)}$ of the self-energies represent the
broadening of these energy levels. All the information about the
ring-to-electrode coupling are included into these two self-energies.

The current passing through the ring is depicted as a single-electron
scattering process between the two reservoirs of charge carriers. The
current $I$ can be computed as a function of the applied bias voltage 
$V$ by the expression,$^{18}$
\begin{equation}
I(V)=\frac{e}{\pi \hbar}\int \limits_{-\infty}^{\infty} 
\left(f_S-f_D\right) T(E) dE
\label{equ8}
\end{equation}
where $f_{S(D)}=f\left(E-\mu_{S(D)}\right)$ gives the Fermi distribution
function with the electrochemical potentials $\mu_{S(D)}=E_F\pm eV/2$. 
Here we assume that the entire voltage is dropped across the ring-electrode 
interfaces, and it is examined that under such an assumption the $I$-$V$ 
characteristics do not change their qualitative features.$^{20}$

All the results in this communication are done at absolute zero temperature, 
but they should valid even for finite temperatures ($\sim 300$ K), since 
the broadening of the energy levels of the ring due to its coupling to the 
electrodes becomes much larger than that of the thermal 
broadening.$^{18}$ On the other hand, at high temperature limit, all 
these phenomena completely disappear. This is due to the fact that the 
phase coherence length decreases significantly with the rise of temperature 
where the contribution comes mainly from the scattering on phonons, and 
accordingly, the quantum interference effect vanishes. For simplicity, we 
take the unit $c=e=h=1$ in our present calculation.

\section{Results and discussion}

To illustrate the results, let us first mention the values of the different 
parameters those are used for the numerical calculation. The on-site 
energy $\epsilon_{i0}$ of the ring is taken as $0$ for all the sites $i$, 
except the sites $i=a$, $b$ and $\alpha$ where the site energies are taken 
as $V_a$, $V_b$ and $V_{\alpha}$, respectively, and the nearest-neighbor 
hopping strength $t$ is set to $3$. On the other hand, for the side 
attached electrodes the on-site energy ($\epsilon_0$) and the 
nearest-neighbor hopping strength ($t_0$) are fixed to $0$ and $4$, 
respectively. The voltage $V_{\alpha}$ is set to $2$, and the Fermi energy
$E_F$ is fixed at $0$ for the whole computation. Throughout the study, we 
focus our results for the two limiting cases depending on the strength 
of the coupling of the ring to the side attached electrodes. In one case 
we use the condition $\tau_{S(D)} << t$, which is the so-called 
weak-coupling limit. For this regime we choose $\tau_S=\tau_D=0.5$. In 
the other case the condition $\tau_{S(D)} \sim t$ is used, which is named 
as the strong-coupling limit. In this particular regime, the values of the 
parameters are set as $\tau_S=\tau_D=2.5$. The significant parameter 
for all these calculations is the AB flux $\phi$ which is set to 
$\phi_0/2$ i.e., $0.5$ in our chosen unit.

\begin{figure}[ht]
{\centering \resizebox*{8cm}{7cm}{\includegraphics{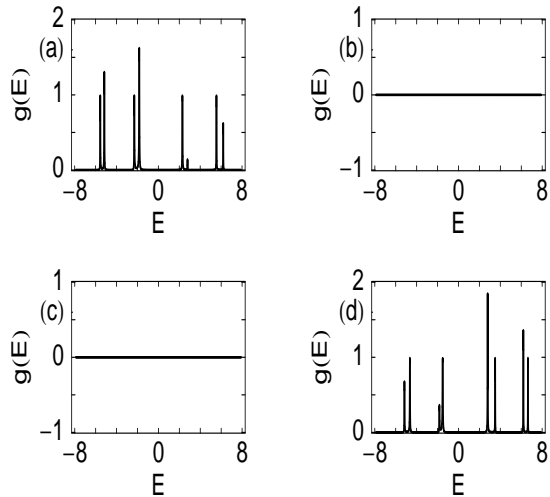}}\par}
\caption{$g$-$E$ curves in the weak-coupling limit for a mesoscopic ring 
with $N=8$, $V_{\alpha}=2$ and $\phi=0.5$. (a) $V_a=V_b=0$, (b) $V_a=2$ 
and $V_b=0$, (c) $V_a=0$ and $V_b=2$ and (d) $V_a=V_b=2$.}
\label{condlow}
\end{figure}

As illustrative examples, in Fig.~\ref{condlow} we describe the 
conductance-energy ($g$-$E$) characteristics for a mesoscopic ring
with $N=8$ and $V_{\alpha}=2$ in the limit of weak-coupling, where (a),
(b), (c) and (d) correspond to the results for the different cases
of the input voltages, $V_a$ and $V_b$. For the particular cases where
anyone of the two inputs is high and other is low i.e., both the two
inputs are not same, the conductance $g$ becomes exactly zero
(Figs.~\ref{condlow}(b) and (c)) for the whole energy range. This
predicts that for these two cases, electron cannot conduct through the 
ring. The situation becomes completely different for the cases when
both the inputs to the gate are the same, either low ($V_a=V_b=0$) or high
($V_a=V_b=2$). In these two cases, the conductance shows fine resonant
peaks for some particular energies (Figs.~\ref{condlow}(a) and (d)), 
which reveal the electron conduction across the ring. At the resonant
energies, $g$ does not get the value $2$, and therefore, the transmission
probability $T$ becomes less than unity, since the expression $g=2T$
is satisfied from the Landauer conductance formula (see Eq.~\ref{equ1}
with $e=h=1$). This reduction of the transmission amplitude is due to the
effect of quantum interference which we will describe below. All these 
resonant peaks are associated with the energy eigenvalues of the ring, 
and thus, it can be predicted that the conductance spectrum manifests 
itself the electronic structure of the ring. Hence, more resonant peaks 
are expected for the larger rings, associated with their energy spectra.
Now we discuss the effects of the gate voltages on the electron transport 
for these four different cases of the input voltages. The transmission 
probability of getting an electron across the ring depends on the quantum 
interference of the electronic waves passing through the two arms (upper 
and lower) of the ring. For the symmetrically connected ring i.e., when 
the two arms of the ring are identical with each other, the probability 
\begin{figure}[ht]
{\centering \resizebox*{8cm}{7cm}{\includegraphics{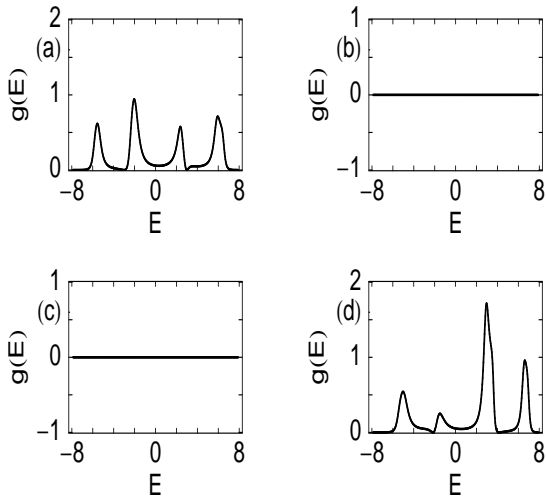}}\par}
\caption{$g$-$E$ curves in the strong-coupling limit for a mesoscopic 
ring with $N=8$, $V_{\alpha}=2$ and $\phi=0.5$. (a) $V_a=V_b=0$, 
(b) $V_a=2$ and $V_b=0$, (c) $V_a=0$ and $V_b=2$ and (d) $V_a=V_b=2$.}
\label{condhigh}
\end{figure}
amplitude becomes exactly zero ($T=0$) for the typical flux, $\phi=\phi_0/2$. 
This is due to the result of the quantum interference among the two waves 
in the two arms of the ring, which can be established through few simple 
mathematical steps. Thus, for the cases when anyone of the two 
inputs to the gate is identical to $2$ and other one is $0$, the upper 
and lower arms of the ring become exactly similar. This is because the 
potential $V_{\alpha}$ is also set to $2$. Accordingly, the transmission 
probability $T$ drops to zero. If the high value ($2$) of anyone of the 
two gates is different from the potential applied in the atomic site 
$\alpha$, then the two arms are not identical with each other and the 
transmission probability will not vanish. Thus, to get the zero 
transmission probability when $V_a$ is high and $V_b$ is low and vice 
versa, we should tune $V_{\alpha}$ properly, observing the input potential.
On the other hand, due to the breaking of the symmetry of the two arms,
the non-zero value of the transmission probability is achieved in the
particular cases when both the two inputs to the gate are the same, which 
reveals the electron conduction across the ring. From these results we 
can emphasize that the electron conduction through the ring takes place 
if both the two inputs to the gate are the same (low or high), while if 
one but not both inputs are high, the conduction is no longer possible. 
This aspect clearly describes the XNOR gate behavior. In addition to these
properties here we also mention the effect of the ring-to-electrode 
coupling. As illustrative examples, in Fig.~\ref{condhigh} we plot the 
$g$-$E$ characteristics for the strong-coupling limit, where (a), (b), 
(c) and (d) are drawn respectively for the same input voltages as in 
Fig.~\ref{condlow}. In the strong-coupling limit, all the resonant peaks 
\begin{figure}[ht]
{\centering \resizebox*{8cm}{7cm}{\includegraphics{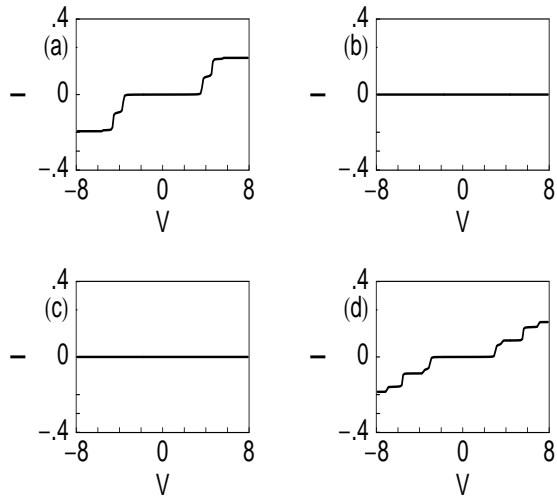}}\par}
\caption{Current $I$ as a function of the bias voltage $V$ for a 
mesoscopic ring with $N=8$, $V_{\alpha}=2$ and $\phi=0.5$ in the limit 
of weak-coupling. (a) $V_a=V_b=0$, (b) $V_a=2$ and $V_b=0$, (c) $V_a=0$ 
and $V_b=2$ and (d) $V_a=V_b=2$.}
\label{currlow}
\end{figure}
get substantial widths compared to the weak-coupling limit. This is 
due to the broadening of the energy levels of the ring in the limit of 
strong-coupling, where the contribution comes from the imaginary parts 
of the self-energies $\Sigma_S$ and $\Sigma_D$, respectively.$^{18}$ 
Therefore, by tuning the coupling strength, we can get the electron 
transmission across the ring for the wider range of energies and it 
provides an important behavior in the study of current-voltage ($I$-$V$)
characteristics. The effects of the gate voltages, $V_a$ and $V_b$, on 
the electron transport in the strong-coupling limit remain exactly similar 
as in the case of weak-coupling limit. 

All these features of electron transport become much more clearly
illustrated from our presented current-voltage ($I$-$V$) characteristics.
The current passing through the ring is determined by integrating the 
transmission function $T$ as prescribed in Eq.~\ref{equ8}. The transmission 
function varies exactly similar to that of the conductance spectrum, 
differ only in magnitude by the factor $2$ since the relation $g=2T$ 
holds from the Landauer conductance formula, Eq.~\ref{equ1}.
As representative examples, in Fig.~\ref{currlow} we plot the $I$-$V$
characteristics for a mesoscopic ring with $N=8$ and $V_{\alpha}=2$,
in the limit of weak-coupling, where (a), (b), (c) and (d) represent
the results for the four different cases of the two input voltages.
From these characteristics it is clearly observed that for the cases when 
one input is high and other is low, the current $I$ becomes exactly zero 
(see Figs.~\ref{currlow}(b) and (c)) for the 
entire bias voltage $V$. This phenomenon can be understood from the 
\begin{table}[ht]
\begin{center}
\caption{XNOR gate behavior in the limit of weak-coupling. The current
$I$ is computed at the bias voltage $6.02$.}
\label{table1}
~\\
\begin{tabular}{|c|c|c|}
\hline \hline
Input-I ($V_a$) & Input-II ($V_b$) & Current ($I$) \\ \hline
$0$ & $0$ & $0.194$ \\ \hline
$2$ & $0$ & $0$ \\ \hline
$0$ & $2$ & $0$ \\ \hline
$2$ & $2$ & $0.157$ \\ \hline \hline
\end{tabular}
\end{center}
\end{table}
conductance spectra, Figs.~\ref{condlow}(b) and (c), since the current 
is computed from the integration method of the transmission function $T$.
The non-zero value of the current appears only when both the two inputs
are identical to zero (Fig.~\ref{currlow}(a)) or high 
(Fig.~\ref{currlow}(d)). The current exhibits staircase-like structure 
with fine steps as a function of the applied bias voltage. This is due 
to the existence of the sharp resonant peaks in the conductance spectrum 
in the weak-coupling limit, since the current is computed by the 
integration method of the transmission function $T$. With the increase
of the bias voltage $V$, the electrochemical potentials on the electrodes
are shifted gradually, and finally cross one of the quantized energy
levels of the ring. Accordingly, a current channel is opened up which
provides a jump in the $I$-$V$ characteristic curve. Here, it is also
important to note that the non-zero value of the current appears beyond a
finite value of the bias voltage $V$, the so-called threshold voltage
($V_{th}$). This $V_{th}$ can be changed by controlling the size ($N$)
of the ring. From these current-voltage spectra, the XNOR gate behavior 
of the ring can be observed very nicely. To make it much more clearer, 
in Table~\ref{table1}, we show a quantitative estimate of the 
typical current amplitude determined at the bias voltage $V=6.02$. It 
shows that, when both the two inputs to the gate are zero, the current 
gets the value $0.194$. While, it ($I$) achieves $0.157$ for the case
when the two inputs are identical to $2$. On the other hand, for the other
two cases, the current becomes exactly zero. In the same fashion, as above, 
\begin{figure}[ht]
{\centering \resizebox*{8cm}{7cm}{\includegraphics{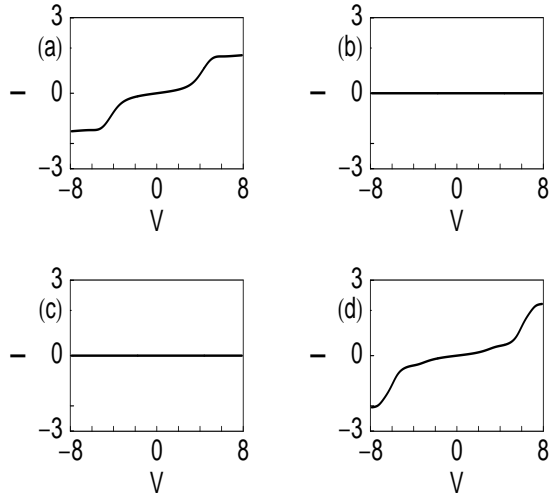}}\par}
\caption{Current $I$ as a function of the bias voltage $V$ for a 
mesoscopic ring with $N=8$, $V_{\alpha}=2$ and $\phi=0.5$ in the limit 
of strong-coupling. (a) $V_a=V_b=0$, (b) $V_a=2$ and $V_b=0$, 
(c) $V_a=0$ and $V_b=2$ and (d) $V_a=V_b=2$.}
\label{currhigh}
\end{figure}
here we also describe the $I$-$V$ characteristics for the limit of
strong-coupling. In this limit, the current varies almost continuously 
with the applied bias voltage and gets much larger amplitude than the 
weak-coupling case (Fig.~\ref{currlow}), as presented in Fig.~\ref{currhigh}. 
The fact is that, in the limit of strong-coupling all the resonant peaks 
get broadened which provide larger current in the integration procedure 
of the transmission function $T$. Thus by tuning the strength of the
ring-to-electrode coupling, we can achieve very large current amplitude
from the very low one for the same bias voltage $V$. All the other
properties i.e., the dependences of the gate voltages on the $I$-$V$
characteristics are exactly similar to those as given in Fig.~\ref{currlow}.
In this strong-coupling limit we also make a quantitative study for the
typical current amplitude, given in Table~\ref{table2}, where the current
amplitude is determined at the same bias voltage ($V=6.02$) as earlier. 
The response of the output current is exactly similar to that as given 
in Table~\ref{table1}. Here the current gets the value $1.466$ for the 
case when $V_a=V_b=0$, and, it goes to $1.174$ when $V_a=V_b=2$. While,
the current exactly vanishes for the other two cases. The non-zero values 
\begin{table}[ht]
\begin{center}
\caption{XNOR gate behavior in the limit of strong-coupling. The current
$I$ is computed at the bias voltage $6.02$.}
\label{table2}
~\\
\begin{tabular}{|c|c|c|}
\hline \hline
Input-I ($V_a$) & Input-II ($V_b$) & Current ($I$) \\ \hline
$0$ & $0$ & $1.466$ \\ \hline
$2$ & $0$ & $0$ \\ \hline
$0$ & $2$ & $0$ \\ \hline
$2$ & $2$ & $1.174$ \\ \hline \hline
\end{tabular}
\end{center}
\end{table}
of the current in this strong-coupling limit are much larger than the 
weak-coupling case, as expected. From these results we can clearly 
manifest that a mesoscopic ring exhibits the XNOR gate response.

\section{Concluding remarks}

In this presentation, we have discussed the XNOR gate behavior of a 
mesoscopic metallic ring threaded by a magnetic flux $\phi$ in the Green's
function formalism. The ring is attached symmetrically to the electrodes
and two gate voltages $V_a$ and $V_b$ are applied in one arm of the ring 
which are taken as the inputs of the XNOR gate. A simple tight-binding 
model is used to describe the system and all the calculations are done 
numerically. We have computed the conductance-energy and current-voltage 
characteristics as functions of the gate voltages, ring-to-electrode 
coupling strength and magnetic flux. Very interestingly we have observed 
that, for the half flux-quantum value of $\phi$ ($\phi=\phi_0/2$), a high 
output current ($1$) (in the logical sense) appears if both the two inputs 
to the gate are the same, either low ($0$) or high ($1$). While, if anyone
of the two inputs is high ($1$) and other is low ($0$), a low output 
current ($0$) results. It clearly manifests the XNOR gate behavior and 
this aspect may be utilized in designing a tailor made electronic logic gate. 

Throughout our study, we have addressed the conductance-energy and
current-voltage characteristics for some fixed parameter values 
considering a ring with total number of atomic sites $N=8$. Though the
results presented here change numerically for the other parameter values 
and ring size ($N$), but all the basic features remain exactly invariant.

In the present work we have done all the calculations by ignoring
the effects of the temperature, electron-electron correlation, disorder,
etc. Due to these factors, any scattering process that appears in the
arms of the rings would have influence on electronic phases, and, in
consequences can disturb the quantum interference effects. Here we
have assumed that in our sample all these effects are too small, and
accordingly, we have neglected all these factors in this particular
study. Beside these factors, in the present study we have also ignored 
the effect of Al'tshuler, Aronov and Spivak (AAS) oscillation. The 
AAS oscillation will appear when either a large amount of fluctuations 
is introduced in the site energies of the ring or the electronic mean 
free path becomes much smaller than the ring size. But, in our model, 
since there is no such fluctuations and also the electronic mean free 
path is comparable to the system size, the effect of AAS oscillation 
will no longer be observed.

The importance of this presentation is mainly concerned with (i) the 
simplicity of the geometry and (ii) the smallness of the size. To the 
best of our knowledge the XNOR gate response in such a simple 
low-dimensional system that can be operated even at finite temperatures
($\sim 300$ K) has not been addressed earlier in the literature.

\vskip 0.3in
\noindent
{\bf\Large Acknowledgment}
\vskip 0.2in
\noindent
I am grateful to the Physical Society of Japan for financial support 
in publication.

\end{document}